# Are digital natives spreadsheet natives?

Maria Csernoch, Piroska Biró
University of Debrecen, Debrecen, Hungary Kassai út 26. 4028
csernoch.maria@inf.unideb.hu, biro.piroska@inf.unideb.hu

**ABSTRACT**

*The present paper reports the results of testing first year students of Informatics on their algorithmic skills and knowledge transfer abilities in spreadsheet environments. The selection of students plays a crucial role in the project. On the one hand, they have officially finished their spreadsheet training – they know everything –, while on the other hand, they do not need any training, since they are digital natives, to whom digital skills are assigned by birth. However, we found that the students had serious difficulties in solving the spreadsheet problems presented; so low were their results that it allowed us to form broad tendencies. Considering computational thinking, algorithmic skills, and knowledge transfer abilities, it is clear that those students performed better who used algorithm-based, multilevel array formulas instead of problem specific, unconnected built-in functions. Furthermore, we can conclude that students, regardless of their birth date and digital generation assigned to them, are in great need of official, high-mathability, algorithm-based training with expert teachers.*

## 1. INTRODUCTION

In the present paper we report the results of testing first year students of Informatics on their algorithmic skills, computational thinking, and knowledge transfer abilities in spreadsheet environments; in general, on their high-mathability problem-solving abilities (Biró & Csernoch, 2015a, 2015b). Furthermore, we questioned how the time spent on digital devices effects the skills and abilities mentioned and the students' problem-solving strategies.

The introductory section outlines the heated debate about the skills and abilities of digital natives, the age group which is our concern in this paper. Following this the aims of our research are outlined, focusing on the future professionals of Informatics in spreadsheet environments. In the Sample section we provide some background information on the tested students relevant to the present paper, the tasks, and the skills required to solve the problems. The Methods section summarizes the testing methods applied and provides the details of the evaluation process. Finally, the Results and Conclusion sections detail the student' achievement in the test, how the digital devices influence their problem-solving abilities, and based on these results guidance for the effective training of 'digital natives' are discussed.

### 1.1. Debates

A heated debate seems to have developed between believers and non-believers around the existence of the species known as 'digital natives'(Prensky, 2001) (Kirschner & De Bruyckere, 2017).The phenomenon of digital native was introduced by Prensky in his famous speech (Prensky, 2001), when he declared that being born after 1984 "endowed this growing group with specific and even unique characteristics that make its members completely different from those growing up in previous generations." (Kirschner & De Bruyckere, 2017). However, this conceptualization – attributed to the huge software companies – derives from the assumed privilege enjoyed by the species of end-users– i.e.

users of "user-friendly" environments –, or "user-friendly-users" for short. In this context the software companies claim that being exposed to graphical interfacesand their selection of tools are sufficient to ensure effective computer problem-solving.

However, it is obvious that neither digital natives nor user-friendly-users can live up to their assumed characteristics. As regards spreadsheets, the European Spreadsheet Risk Interest Group (EuSpRIG) is devoted to revealing the consequences of the activities of untrained users and to finding solutions for handling the problems of error-prone documents (Horror Stories, 2017). In accordance with these efforts EuSpRIG introduced the "Twenty principles for good spreadsheet practice" in 2015 (TPDSP, 2015), the "Spreadsheet competency framework" in 2016 (SCF, 2016), and accepted and published the paper entitled "Edu-Edition Spreadsheet Competency Framework" in 2017 (Csernoch & Biró, 2018). All these interconnected research studies and documents focus on training user-friendly-users to understand better, to develop their computational thinking, to improve their problem-solving skills, and, in general, to make them more productive.

## 2. Aims

In this research, our aim was to measure with quantitative tools and methods how students of Informatics relate to spreadsheets problems, and how they transfer fundamental algorithms between problems and environments. We wanted to find connections between the students' problem-solving and knowledge-transfer abilities (the sample is detailed in Section 3) in spreadsheet environments, considering both internal and external sources of knowledge and information. In close connection with this aim, we also were interested in finding out how students understand and build algorithms in this unofficial programming environment.

Our other aim was to measure how computer and mobile-device-time – in the present context this means mobile devices functioning as computers –influence students' performance in understanding spreadsheet problems, building algorithms, and coding them with spreadsheet formulas.

## 3. SAMPLE

### 3.1. Three Groups of Students of Informatics

Three groups of students of Informatics, during their first week of tertiary studies were tested at the University of Debrecen in the academic year of 2016/2017 as part of the TAaAS (Testing Algorithmic and Application Skills) project launched in 2011 (Csernoch et al., 2015). The three majors taught at the faculty made up the three groups: Software Engineering, System Engineering, and Business Informatics (SOE, SYE, and BIM, respectively) (for details of the majors see Csernoch et al., 2015).

As regards the students' background knowledge, we know that they finished their high school studies, where Informatics is a compulsory subject with a varying – and untraceable –number of classes (Kerettanterv, 2013), that most of the students passed the maturation exams in Informatics at either the intermediate or/and advanced levels (Érettségi vizsga, 2017), with no significant differences between the three groups (Csernoch et al, 2015), and that many of them have passed the ECDL exams (ECDL Hungary, 2017).

In general, we were testing the future professionals of Informatics and Computer Sciences, who finished their official training in spreadsheets and other birotical environments; in short, the (future) professionals of the subject.

### 3.2. Tasks

The spreadsheet tasks involved in the research were presented in the form of

- a table with five fields of data and 235 records (Figure 1),
- five open questions to answer with spreadsheet formulas (Figure 2), and
- a spreadsheet array formula to decode, to answer with a complete natural language sentence (Figure 3).

| | A | B | C | D | E | F | G |
|---|---|---|---|---|---|---|---|
| 1 | Country | Continent | Capital | Area | Population (thousand) | | |
| 2 | Afghanistan | Asia | Kabul | 647500 | 27756 | | |
| 3 | Albania | Europe | Tirana | 28748 | 3545 | | |
| 4 | Algeria | Africa | Algiers | 2381740 | 32278 | | |
| 5 | American Samoa | Oceania | Pago Pago | 199 | 69 | | |
| 6 | Andorra | Europe | Andorra la Vella | 468 | 68 | | |
| 7 | Angola | Africa | Luanda | 1246700 | 10593 | | |
| 8 | Anguilla | America | The Valley | 102 | 12 | | |
| 233 | Yemen | Asia | Sanaa | 527970 | 18701 | | |
| 234 | Yugoslavia | Europe | Belgrade | 102350 | 10657 | | |
| 235 | Zambia | Africa | Lusaka | 752614 | 9959 | | |
| 236 | Zimbabwe | Africa | Harare | 390580 | 11377 | | |

*Figure 1. The sample table of the TAaAS spreadsheet test with the shaded G2 cell referring to the irrelevance of the value stored in it.*

a) What is the capital city of the largest country?

b) What is the population density of each country?

c) How many African countries are in the table?

d) What is the average area of those countries whose population is smaller than G2?

e) How many countries have a surface area greater than G2?

*Figure 2. Questions of the test to be answered with spreadsheet formulas.*

f) What is the result of the following formula?
   {=SUM(IF(B2:B236="Europe",IF(LEFT(A2:A236)="A",1)))}

*Figure 3. The decoding task of the test to be answered with a natural language sentence.*

### 3.3. The Skills Required to Solve the Presented Problems

**Internal skills and knowledge**

- Information retrieval from the table:
  - A thorough data analysis of the table presented in Figure 1. In this case the data source is the table and no further external knowledge is required.
  - Recognizing that the population is stored in thousands.

- Information retrieval from the tasks:
    - Recognizing the input values of the tasks (Figure 1 and Figure 2).
    - Recognizing the output values of the tasks (Figure 1and Figure 2).
- Recognizing the G2 cell as a variable(Figure 1 and Figure 2).
- Recognizing the same algorithm in Tasks a–f(Figure 2 and Figure 3).
- Decoding (reading) the presented formula in Task f (Figure 3).

**External skills and knowledge**

- Understanding the expression "using a formula" to solve problems.
- Understanding the difference between formula-output and constant-output.
- Handling variables in spreadsheets.
- Understanding the difference between a constant and a variable.
- Task a:
    - Building or recalling the algorithm of the capital city problem.
    - Recalling the functions and their arguments to solve the problem.
    - Building multilevel functions.
- Task b:
    - How to calculate population density: building or recalling the algorithm.
    - How to expand the population density of one country: recalling how to a copy formula or create an array formula.
- Tasks c–e, solved with array formulas (AF):
    - Applying the algorithm of conditional summing and its modification according to the criteria.
    - Recalling the functions and their arguments to solve the problem (SUM(), AVERAGE(), IF()).
    - Building multilevel functions.
    - Recognizing that Task e is a two-step-generalization of Task c (inequality and variable).
    - Recognizing that Task d is a three-step-generalization of Task c (inequality, variable, and average instead of sum).
- Task c, solved with build-in formulas (BIF):

- Recalling at least one of the suitable functions and its arguments (COUNTIF() / COUNTIFS() / DCOUNT() / DCOUNTA() / DGET()).

– Task d, solved with build-in formulas:

- Recalling at least one of the suitable functions and its arguments (AVERAGEIF() / AVERAGEIFS() / DAVERAGE() / DGET()).

- Recalling the syntax of handling inequality with variables.

– Task e, solved with build-in formulas:

- Recalling at least one of the suitable functions and its arguments (COUNTIF() / COUNTIFS() / DCOUNT() / DCOUNTA() / DGET()).

- Recalling the syntax of handling inequality with variables in the case of the *IF?() functions.

- Recalling the database functions (DBF).

- Creating the conditional grid to the database functions.

– Tasks f:

- Recognizing the functions and their arguments (SUM(), IF(), LEFT()).

- Recognizing multilevel formulas.

- Understanding array formulas.

### 3.4. Methods

To see how the length of time students spend on computers and on mobile devices functioning as computers influences their problem-solving abilities and algorithmic skills in spreadsheets, the periods of time were recorded in the attitude test of the TAaAS project. For both sets of devices three options were offered: at least 5 hours every day (c5 and m5), at least 2 hours every day (c2 and m2), and less than 2 hours every day (c1 and m1), where c stands for computers and m for mobile devices. Based on these categories, we referred to the students as heavy, moderate, and occasional users, respectively.

*Table 1. The number of students participating in the project and the time they spend on computers and mobile devices (functioning as computers).*

|     | N   | c5  | c2  | c1  | m5  | m2  | m1  |
|-----|-----|-----|-----|-----|-----|-----|-----|
| SOE | 120 | 50  | 62  | 7   | 27  | 50  | 41  |
| SYE | 103 | 47  | 51  | 5   | 23  | 52  | 28  |
| BIM | 97  | 32  | 52  | 11  | 42  | 36  | 19  |
|     | 320 | 129 | 165 | 23  | 92  | 138 | 88  |

*Table 2. The number of recognizable items of the TAaAS spreadsheet problems (BIF and AF refer to built-in functions and array formulas, respectively).*

| Task a | Task b | Task c | | Task d | | Task e | | Task f |
|---|---|---|---|---|---|---|---|---|
| | | BIF | AF | BIF | AF | BIF | AF | |
| 15 | 4 | 6 | 10 | 10 | 9 | 8 | 9 | 3 |

Considering the students' algorithm skills and knowledge transfer abilities it was not only the syntactically correct answers which were accepted, but the all identifiable algorithms, regardless of the language used. Any natural language algorithm or pseudo-code was considered and evaluated according to the predefined items.

In cases where either built-in functions or array formulas are acceptable the items were set up according to the skills and knowledge required in the specific cases (Table 2). We must note here that in the TAaAS project, in general, an extremely low number of students selected database functions (DBF), while in the three groups in 2016 none of the students tried them. Consequently, in the present paper the BIF abbreviation refers to the *IF?() functions.

The numbers of recognizable items of Tasks c–e clearly show that applying the array formulas requires the application of the same number of items, while using the built-in functions requires various numbers of items. (In Task c, the number of AF items is 10, compared to 9 in Tasks d and e, due to the string constant in the formula.)

## 4. RESULTS

The students' scores were calculated in two different ways. First the correct answers were collected, where their numbers and average numbers were calculated. Table 3 clearly indicates that we can hardly find students who were able to complete these tasks. This method was then further tuned. In the following process the solutions were evaluated according to the predefined items, and the number of the correct items was calculated and used for the wider analyses.

*Table 3. The percentage of the students who solved the problems completely.*

|  | Task a | Task b | Task c | Task d | Task e | Task f |
|---|---|---|---|---|---|---|
| SOE | 3.33% | 1.67% | 15.83% | 1.67% | 2.50% | 25.00% |
| SYE | 0.00% | 0.97% | 9.71% | 0.00% | 0.97% | 27.18% |
| BIM | 0.00% | 0.00% | 8.25% | 2.06% | 2.06% | 14.43% |

*Table 4. The average results of the students based on the predefined items.*

|  | Task a | Task b | Task c | Task d | Task e | Task f |
|---|---|---|---|---|---|---|
| SOE | 21.06% | 25.21% | 35.92% | 15.53% | 15.23% | 31.94% |
| SYE | 16.12% | 25.49% | 42.20% | 15.06% | 15.22% | 37.86% |
| BIM | 16.56% | 25.52% | 33.40% | 18.28% | 14.30% | 23.37% |

It is clear from the result tables (Table 3 and Table 4) that students have difficulties in solving these problems, considering the skills required in the test. As is detailed in the list of skills and knowledge, there are fundamentally three approaches to solve Tasks c–e: "hand-made" array formulas, built-in *IF?()functions, or built-in DBF() functions. Aware of these possible solutions and the students' selections, we also compared the number and the results of the students selecting the BIF or the AF solutions.

### 4.1. Knowledge transfer

*Table 5. The number of students selecting the BIF (*IF?() functions) or the AF solutions in Tasks c–e.*

|        | Task c |     | Task d |     | Task e |     |
|--------|--------|-----|--------|-----|--------|-----|
|        | BIF    | AF  | BIF    | AF  | BIF    | AF  |
| SOE    | 47     | 24  | 14     | 33  | 20     | 21  |
| SYE    | 40     | 23  | 7      | 26  | 20     | 12  |
| BIM    | 33     | 17  | 9      | 28  | 11     | 14  |
|        | 120    | 64  | 30     | 87  | 51     | 47  |
| BIF/AF | 1.875  |     | 0.3448 |     | 1.0851 |     |

When we compare the usage of BIF and AF in Tasks c–e, which apply the same algorithm, it is clear from the data of Table 5 that the students' choice of solution is highly determined by the level of generalization. In Task c, where equivalence is checked in the condition (which can be omitted in the BIF solution) and a string constant is used, the BIF/AF rate is the highest. Furthermore, we found that the higher the generalization level, the lower the BIF/AF rate: Task c > Task e > Task d (1.875, 1.0851, and 0.3448, respectively). This finding indicates that students apply the BIF solution only in simple problems and cannot expand this knowledge to more demanding problems.

Considering the average results of the students who worked with the problems, regardless of the students' selection of methods – i.e. regardless of the BIF/AF rate – and regardless of the generalization level of the problem, the results of the AF solutions are found to be higher than those of the BIF solutions (Table 6) in all the three groups.

*Table 6. The results of the students selecting BIF or AF solutions in Tasks c–e. Since the choice made was tested at this stage of the analysis, this table shows only the results of those who tried the tasks (Table 5).*

|     | Task c  |         | Task d  |         | Task e  |         |
|-----|---------|---------|---------|---------|---------|---------|
|     | BIF     | AF      | BIF     | AF      | BIF     | AF      |
| SOE | 57.45%  | 67.08%  | 30.71%  | 43.43%  | 42.50%  | 47.62%  |
| SYE | 66.67%  | 73.04%  | 34.29%  | 50.43%  | 40.63%  | 62.96%  |
| BIM | 63.64%  | 67.06%  | 37.78%  | 51.19%  | 44.32%  | 64.29%  |
|     | 62.59%  | 69.06%  | 34.26%  | 48.35%  | 42.48%  | 58.29%  |

Focusing on knowledge transfer abilities, we checked the correlation between the a-c BIF and between the a–c AF solutions. At this point we must note that the extremely low number of students who tried to solve the problems (Table 5) and their low average results (Table 3 and Table 4) requires further statistical analyses on different samples; however, the tendencies are clear and require further attention.

The correlation between the BIF solutions (Table 7) and the AF solutions (Table 8) were calculated and correlation matrixes were built based on the intervals of the weak (W), moderate (M), and strong (S) connections ([0, 0.3), [0.3, 0.7), and [0.7, 1], respectively).

*Table 7. The correlation matrix of the BIF solutions in Tasks c–e.*

| BIF | Students who tried (ST) |     |     | All students (SA) |     |     |
|-----|-----|-----|-----|-----|-----|-----|
|     | SOE | SYE | BIM | SOE | SYE | BIM |
| C-D | W   | M   | M   | M   | M   | M   |
| C-E | M   | M   | M   | M   | M   | M   |
| D-E | W   | S   | W   | M   | W   | M   |

The correlation matrixes can show tendencies relating to how the students are able to transfer their spreadsheet knowledge between problems sharing the same algorithm. The number of weak, middle and strong correlations were counted both in the BIF (Table 7) and the AF (Table 8) solutions. Since the low number of students who dealt with the problems is not enough to form firm conclusions, we will discuss the tendencies. The rate of the three correlation groups are the following, considering those students who at least tried to do something: 3:5:1 and 1:4:4 in the BIF and the AF groups, respectively. This result indicates that those students who use array formulas can transfer the background knowledge required by this type of coding more effectively than those who use the problem specific built-in functions.

*Table 8. The correlation matrix of the AF solutions in Tasks c–e.*

| AF | Students who tried (ST) | | | All students (SA) | | |
|---|---|---|---|---|---|---|
| | SOE | SYE | BIM | SOE | SYE | BIM |
| C-D | W | M | M | W | M | M |
| C-E | M | S | S | M | W | M |
| D-E | M | S | S | M | M | M |

In the case of decoding Task f, in general, we found no correlation between the BIF and AF users. In this respect the weak, middle, and strong correlations of the three groups show such different patterns that this analysis requires further samples to form conclusions. However, on this sample we had to reject our hypothesis that those who write array formulas can decode them, or vice versa. It is remarkable that even those who used AFs in solving Tasks c–e, were not able to transfer their knowledge to decoding in all the groups– no strong correlation was found between the two activities. At this stage of our analyses we can suggest that this result means that the students are code-, rather than algorithm-dependent, which will further be proved by the time spent on computers (Section 4.3). Another explanation would be that writing codes requires different skills than decoding and forming answers in natural language sentences, and students are trained to write codes rather than read them.

Considering Task b, we found that students have problems utilizing both external and internal knowledge. However, internal knowledge, information retrieved from the table (which is our main concern now) caused more problems than the question of how to calculate the population density. Students were not able to recognize that the population is presented in thousands in the table and that an array formula or one formula with copying should be created. This finding reveals that the students tested are not used to data analysis in spreadsheet environments.

*Table 9. The results of Task a ,considering the items and the order of the three functions of the multilevel function.*

| | Task a | Task a without MAX() | The order of INDEX(MATCH(MAX())) |
|---|---|---|---|
| SOE | 21.06% | 16.35% | 18.33% |
| SYE | 16.12% | 11.65% | 11.65% |
| BIM | 16.56% | 11.42% | 12.37% |

As is shown in Table 3, only an extremely low number of students were able to solve Task a correctly. Considering the items, most of the students recognized that the largest area can be calculated with the MAX() function. However, in general this is all they know. They cannot formulate the algorithm, nor build the multilevel function, and are not aware of the arguments of the INDEX() and the MATCH() functions (Table 9). Several students noted that they could solve the problem with computers, something which on the one hand, we cannot prove, while on the other hand it was not the aim of the present analyses. However, our teaching experience clearly shows that students who have passed either the maturation exams or the ECDL exams are not familiar with the algorithm for finding an item in a vector and writing out the corresponding value from another.

We also must note that even though the table structure ruled out the use of both VLOOKUP() and HLOOKUP() several students selected them in their solutions.

### 4.2. Misconceptions

We cannot ignore the misconceptions revealed, both in our teaching experience and in this test. However, we are glad to report that the number of those who claimed that the formula of Task f is incorrect is significantly lower compared to previous years and compared to the results of the teachers of Informatics, a tendency which is rather promising. In a similar way, in Task a, the number of students who wanted to solve the problem with the BIF HLOOKUP() or VLOOKUP() functions decreased, and in this analysis is below 10%.

At this point, we also must note that the helps available for the MATCH() function, 20 years after MS Excel was published, are still incorrect and inconsistent, stating that the second argument of the MATCH() function can be any range lookup array. Required. The range of cells being searched." (MATCH function, 2017) or an array(Figure 4), instead of a vector(Csernoch, 2017; Csernoch & Biró, 2018). These errors do not help the understanding of the search algorithms and the application of the MATCH() function.

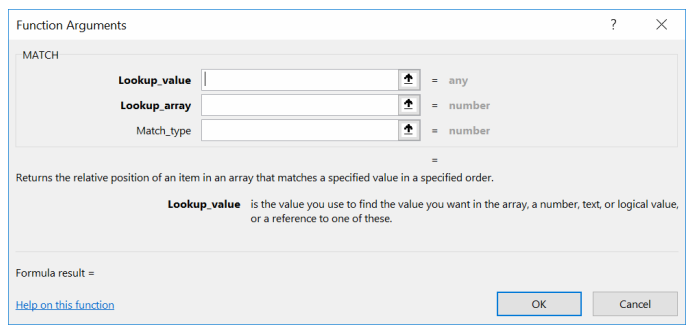

*Figure 4. The wizard of the MATCH() function in Excel2016.*

### 4.3. Time Spent on Computers and Mobile Devices

Our other major concern was how the use of computers and mobile devices influences the skills in question. As mentioned in the Methods section and shown in Table 1, we formed three groups of students based on the time they spend on computers and mobile devices functioning as computers.

In all the groups in all the tasks – BIF and AF solutions are not separated in this step of the analysis – we compared the results. The average results of the six groups – c5, c2, c1 and m5, m2, m1 – were calculated for each task. Based on the average results, computer and mobile use were separated, and for both categories the ranks of the averages were established. Finally, the sums of these ranks were calculated and mapped in 3D column charts.

We ran a competition and applied the method of those sports where the highest point is the best rank. The highest result in the task is the best and marked with the lowest rank. In the diagram, they are the lowest columns (Figure 5 and Figure 6). The three groups are mapped one by one and finally their summed ranked is mapped as 3G.The ranking reveals minor differences in the comparison of the three groups (with no significant differences),so, we can form our conclusions considering the three groups together.

Considering the use of computers, it is clear that they have positive effects on the results of solving the problems presented by using the skills and knowledge tested (Figure 5), however mainly on creating formulas. The heavy computer user students achieved the highest results with the best ranks, while the occasional users scored the lowest with the worst ranks. The BIM group is somewhat different from the other two groups and from the average, since it is stronger in the moderate users than in the heavy users.

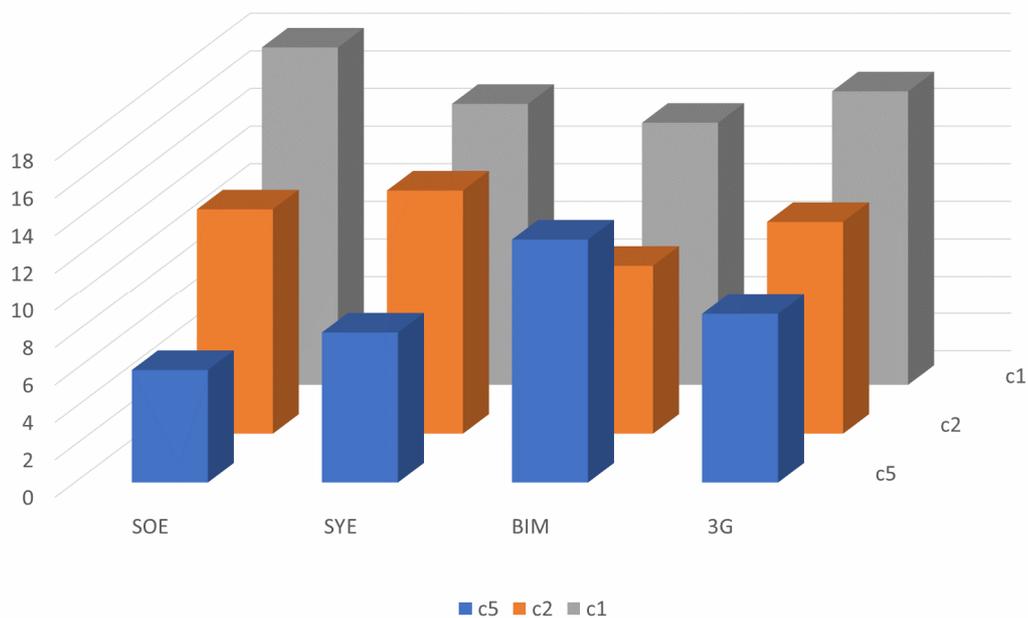

*Figure 5. The summed ranks of the results, in comparison with time spent on computers.*

Our further question was whether 'artificial hands', i.e. mobile devices, have positive effects on the students' spreadsheet skills, or not. Do they need formal training in the subject or do smart mobile devices provide enough tools, help, and aid?

In this regard, we have found that mobile devices do not improve the spreadsheet skills of the students tested. However, the good news is that we have not encountered any negative effects, either (Figure 6). In general, these devices do not help the development of computational thinking, algorithmic skills, and knowledge transfer abilities among this group as was claimed by Prensky (2001). We can conclude that these students need training (Kirschner et al., 2006),guidance from expert teachers (Hattie, 2003, 2012) who believe in the incremental nature of science (Chen et al., 2015), and are well trained in all aspects of TPCK (Technological, Pedagogical, and Content Knowledge, Mishra & Koehler, 2006); these teachers are not mobile devices, since they do not provide sufficient help.

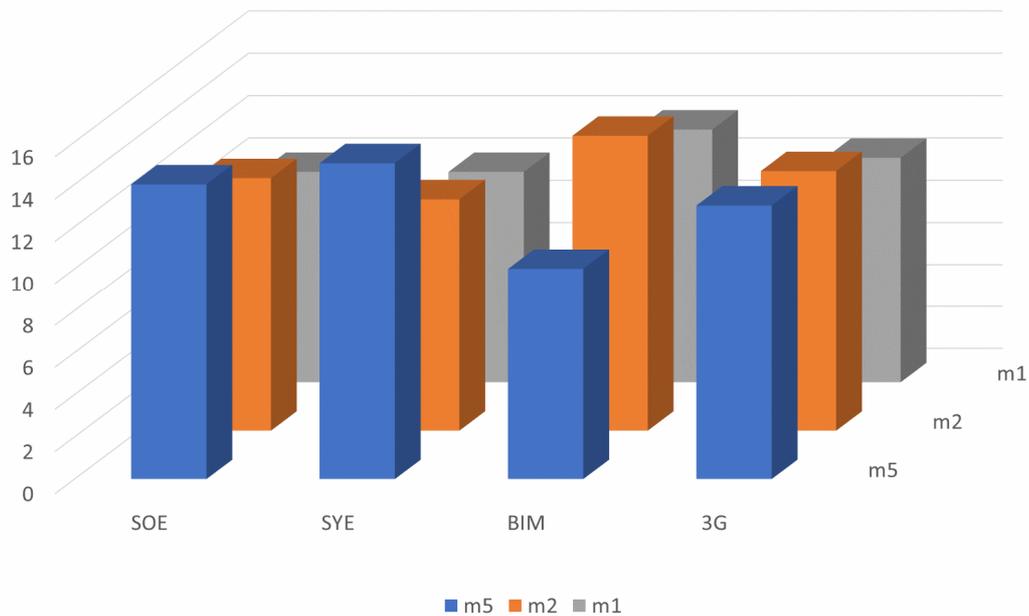

*Figure 6. The summed ranks of the results, in comparison with time spent on mobile devices functioning as computers.*

We also must note that we have to complete further analyses on further samples since the average results of the students were so low that unsolved problems greatly affect the results.

**5. CONCLUSION**

Our analyses reveal the tendencies regarding the algorithm skills and knowledge transfer abilities of the students tested. It was found in the conditional summing tasks that the use of problem specific built-in functions requires mostly unconnected, non-conventional knowledge, and that the application of these functions constitutes non-transferable knowledge. The application of array formulas, however, requires the knowledge or recognition of the shared algorithm and reveals identifiable connections between the three tasks, which is in accordance with the skills preferred in the $E^2SCF$ rather than the original SCF. On the other hand, even the skill of creating array formulas is not enough to decode and read similar formulas.

In a similar way, the results of the students, especially in calculating the population density, proved that information retrieval from tables should be in the focus from the very beginning of spreadsheet studies, as is suggested in the $E^2SCF$. Without this skill, the students' external knowledge cannot be built into spreadsheet formulas.

Considering the effect of the time spent on computers and mobile devices functioning as computers, we have found that computers have a positive effect on the skills and abilities tested, while mobile devices are neutral. These findings do not prove the ideas of Prensky (2001) that digital natives are born with digital devices and born with the skills required for their intelligent use, but rather support the ideas of Kirschner & De Bruyckere (2017) that digital skills have to be taught and gained through well-structured training. This latter finding is in complete accordance with our findings considering the knowledge transfer, algorithmic skills, and computational skills detected in the solutions of our tested students.

TPGSP. (2015), "Twenty principles for good spreadsheet practice". ICAEW. Second edition. Retrieved November 21, 2016 from http://www.icaew.com/-/media/corporate/files/technical/information-technology/excel-community/166-twenty-principles-for-good-spreadsheet-practice.ashx?la=en